\documentclass{article}
\usepackage{ijcai23}

\usepackage{times}
\usepackage{soul}
\usepackage{url}
\usepackage[hidelinks]{hyperref}
\usepackage[utf8]{inputenc}
\usepackage[small]{caption}
\usepackage{graphicx}
\usepackage{amsmath}
\usepackage{amsthm}
\usepackage{booktabs}
\usepackage{algorithm}
\usepackage{algorithmic}
\usepackage[switch]{lineno}

\usepackage{tabularx}
\usepackage{subcaption}
\usepackage{multirow}
\usepackage{amsfonts,amssymb,bbding,pifont,xcolor,makecell}
\newcommand{\cmark}{\ding{51}}%
\title{Cross-Modal Global Interaction and Local Alignment for Audio-Visual Speech Recognition}

\author{
Yuchen Hu$^1$ \and Ruizhe Li$^2$ \and Chen Chen$^1$ \and Heqing Zou$^1$ \and Qiushi Zhu$^{3}$ \And Eng Siong Chng$^1$
\affiliations
$^1$Nanyang Technological University, Singapore
\quad$^2$University of Aberdeen, UK\\
$^3$University of Science and Technology of China, China
}

\begin{document}

\maketitle

\begin{abstract}
Audio-visual speech recognition (AVSR) research has gained a great success recently by improving the noise-robustness of audio-only automatic speech recognition (ASR) with noise-invariant visual information.
However, most existing AVSR approaches simply fuse the audio and visual features by concatenation, without explicit interactions to capture the deep correlations between them, which results in sub-optimal multimodal representations for downstream speech recognition task.
In this paper, we propose a cross-modal global interaction and local alignment (GILA) approach for AVSR, which captures the deep audio-visual (A-V) correlations from both global and local perspectives.
Specifically, we design a global interaction model to capture the A-V complementary relationship on modality level, as well as a local alignment approach to model the A-V temporal consistency on frame level.
Such a holistic view of cross-modal correlations enable better multimodal representations for AVSR.
Experiments on public benchmarks LRS3 and LRS2 show that our GILA outperforms the supervised learning state-of-the-art\footnote{Code is available at \url{https://github.com/YUCHEN005/GILA}.}.


\end{abstract}

\section{Introduction}
With recent advancement of deep learning techniques, automatic speech recognition (ASR) has achieved quite good performance~\cite{graves2012sequence,vaswani2017attention,chen2022self}. However, ASR systems are usually vulnerable to noise and would degrade significantly under noisy conditions~\cite{sumby1954visual}. To improve their performance under various scenarios, recent works on noise-robust speech recognition have made some progress~\cite{wang2020complex,chen2022noise,hu2022interactive,zhu2023robust}.

A currently popular research direction on robustness combines audio (A) and visual (V) features to benefit from the noise-invariant lip movement information. With use of two modalities, audio-visual speech recognition (AVSR) systems move one step closer to how human perceives speech~\cite{sumby1954visual} and achieve better performance in many application scenarios~\cite{biswas2016multiple,koguchi2018mobile}.
Thanks to recent advance of neural network, AVSR has achieved a remarkable success~\cite{afouras2018deep,makino2019recurrent,ma2021end,pan2022leveraging,chen2022leveraging,shi2022robust,hsu2022u,zhu2023vatlm}.
However, most existing AVSR works simply employ feature concatenation for audio-visual (A-V) fusion, without explicit interactions to capture deep correlations between them~\cite{raij2000audiovisual}:
1) From global perspective, they may not capture the complementary relationship between A-V modalities. 
Such relationship means when one modality is missing or corrupted, the other modality can supply valid information for downstream task~\cite{wang2022deep}.
Failure to capture it would make the system confused about the significance of each modality and thus degrade the performance~\cite{hori2017attention,tao2018gating}.
2) From local perspective, they may ignore the temporal alignment between A-V frames, which could be a problem due to the ambiguity of homophenes~\cite{kim2022distinguishing} where same lip shape could produce different sounds.
Such misalignment between lip and audio sequences would increase the difficulty of efficient multimodal fusion and affect final performance ~\cite{tsai2019multimodal,lv2021progressive}.

To capture the global complementary relationship between different modalities, cross-attention has been widely investigated in recent multimodal studies to learn the inter-modal correspondence~\cite{lee2020cross,li2021align,goncalves2022auxformer,mercea2022audio}.
Despite the effectiveness, it fails to simultaneously preserve the intra-modal correspondence that could adaptively select the information of each individual modality for the inter-modal correspondence modeling~\cite{wang2022deep}, which thus results in sub-optimal complementary relationship between modalities.

From the local perspective, contrastive learning has been popular for cross-modal temporal alignment to model the frame-level consistency~\cite{korbar2018cooperative,morgado2021audio,hu2022dual,yang2022vision}, but they seem to only align the multimodal features within same model layer, ignoring the alignment across different layers. 
Since different-layer features contain semantic representations of different granularities~\cite{gu2021image}, we argue that the alignment between them could capture extra contextual information to improve the modeled temporal consistency.


In this paper, we propose a cross-modal global interaction and local alignment (GILA) approach to effectively capture the deep audio-visual correlations from both global and local perspectives.
Specifically, we propose an attention-based global interaction (GI) model to capture the A-V complementary relationship on modality level.
On top of the vanilla cross-attention, we propose a novel iterative refinement module to jointly model the A-V inter- and intra-modal correspondence. 
It could adaptively leverage the information within each individual modality to capture the inter-modal correspondence, which thus results in better complementary relationship between A-V modalities.
With global knowledge of A-V correlations, the system may still be less aware of the local details.
To this end, we further design a cross-modal local alignment (LA) approach via contrastive learning to model the A-V temporal consistency on frame level.
Based on the vanilla within-layer alignment, we propose a novel cross-layer contrastive learning approach to align A-V features across different GI model layers.
Such design could capture extra contextual information between the different-granularity semantic representations, which enables more informative temporal consistency between A-V frames.
As a result, our proposed GILA can capture deep holistic correlations between A-V features and finally generate better multimodal representations for downstream recognition task.

To the best of our knowledge, this is the first AVSR work to model deep A-V correlations from both global and local perspectives. Our main contributions are summarized as:

\begin{itemize}
\item We present GILA, a novel approach to capture deep audio-visual correlations for AVSR task, from both global and local perspectives.

\item We propose a cross-modal global interaction (GI) model to capture A-V complementary relationship on modality level, as well as a local alignment (LA) approach to model the A-V temporal consistency on frame level.

\item Experimental results on two public benchmarks demonstrate the effectiveness of our GILA against the state-of-the-art (SOTA) supervised learning baseline, with up to 16.2\% relative WER improvement.
\end{itemize}

\section{Related Work}
\label{sec:related_work}
\noindent\textbf{Audio-Visual Speech Recognition.}
Most existing AVSR works focus on novel architectures and supervised learning methods, investigating how to effectively model and fuse the audio-visual modalities.
TM-seq2seq~\cite{afouras2018deep} proposes a Transformer-based~\cite{vaswani2017attention} AVSR system with sequence-to-sequence loss.
Hyb-RNN~\cite{petridis2018audio} proposes a RNN-based AVSR system with hybrid seq2seq/CTC loss~\cite{watanabe2017hybrid}.
RNN-T~\cite{makino2019recurrent} employs recurrent neural network transducer~\cite{graves2012sequence} for AVSR task.
EG-seq2seq~\cite{xu2020discriminative} builds a joint audio enhancement and multimodal speech recognition system based on RNN.
LF-MMI TDNN~\cite{yu2020audio} proposes a joint audio-visual speech separation and recognition system based on TDNN.
Hyb-Conformer~\cite{ma2021end} proposes a Conformer-based~\cite{gulati2020conformer} AVSR system with hybrid seq2seq/CTC loss, where the audio-visual streams are encoded separately and then concatenated for decoding, which has achieved the supervised learning SOTA on both LRS3 and LRS2 datasets.
MoCo+wav2vec~\cite{pan2022leveraging} employs self-supervised pre-trained audio/visual front-ends to improve AVSR performance, which has achieved the SOTA on LRS2 dataset.
However, these studies simply concatenate the audio and visual features for multimodal fusion, without explicit interactions to capture their deep correlations.
Recently proposed AV-HuBERT~\cite{shi2022learning,shi2022robust} employs self-supervised learning to capture contextual correlations between audio-visual features, and the latest u-HuBERT~\cite{hsu2022u} extends it to a unified framework of multimodal and unimodal pre-training, which has achieved the SOTA on LRS3 dataset. 
However, they require a large amount of unlabeled data and computing resources.
In this work, we propose a novel supervised learning approach called GILA to efficiently capture deep A-V correlations from both global and local perspectives.

\vspace{0.1cm}
\noindent\textbf{Cross-Modal Modality-Level Interaction.} Attention methods have been widely investigated to interact between different modalities to capture their complementary relationship, in various multimodal applications such as A-V emotion recognition~\cite{goncalves2022auxformer}, A-V action localization~\cite{lee2020cross}, etc. 
Recent works employ cross-attention to enable extracted features of different modalities to attend to each other~\cite{lee2020cross,li2021align,goncalves2022auxformer,mercea2022audio}, which is found effective to capture the inter-modal correspondence and significantly improves the system performance.
However, they may not simultaneously preserve the intra-modal correspondence that could adaptively select the unimodal information for inter-modal correspondence modeling~\cite{wang2022deep}.
To this end, we propose a novel iterative refinement module to jointly model the inter- and intra-modal correspondence, where the key idea is introducing a bottleneck feature to recurrently collect multimodal information.

\vspace{0.1cm}
\noindent\textbf{Cross-Modal Frame-Level Alignment.}
Cross-modal alignment aims to model the temporal consistency between sequences of different modalities, and alleviate the frame-level misalignment problem in some scenarios~\cite{tsai2019multimodal,lv2021progressive,kim2022distinguishing}.
This is typically done by contrastive learning where the correspondence between positive pairs is trained to be stronger than those of negative pairs~\cite{chopra2005learning}.
Recently, contrastive learning is popular for cross-modal temporal alignment, which has achieved significant improvement on various tasks~\cite{korbar2018cooperative,hadji2021representation,morgado2021audio,yang2022vision}.
However, they seem to only align features of multiple modalities within same model layer, ignoring the alignment across different layers that could learn extra contextual information between different-granularity semantic representations.
In this work, we propose a cross-layer contrastive learning approach for holistic A-V alignments.

\begin{figure*}[ht]
\centering
\includegraphics[width=0.93\textwidth]{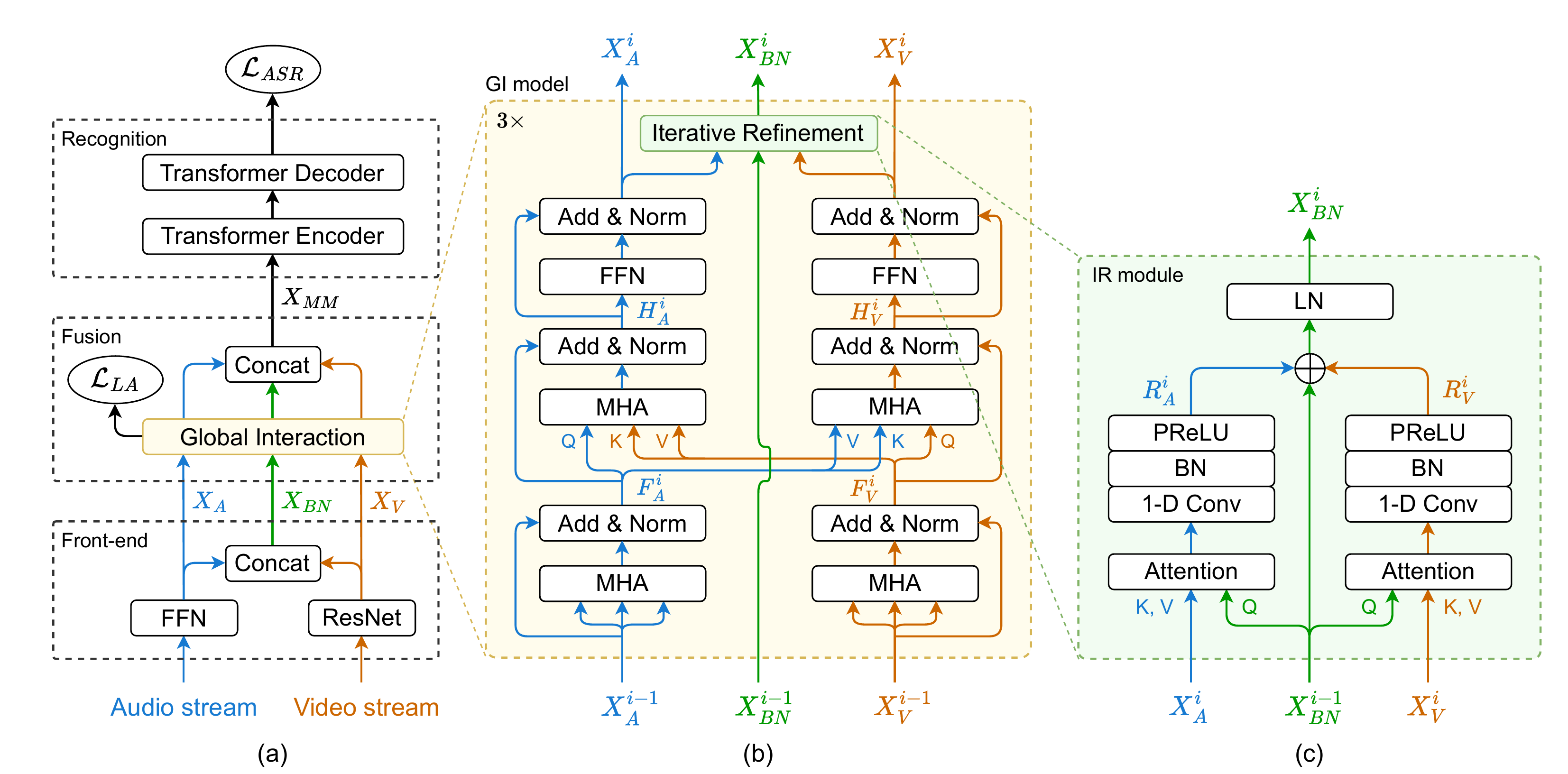}
\vspace{-0.1cm}
\caption{Block diagrams of proposed GILA: (a) Overall architecture, (b) Global Interaction (GI) model, (c) Iterative Refinement module. 
The $\mathcal{L}_{A\hspace{-0.01cm}S\hspace{-0.01cm}R}$ denotes speech recognition loss, and $\mathcal{L}_{L\hspace{-0.01cm}A}$ denotes local alignment loss.}\label{fig1}
\vspace{-0.1cm}
\end{figure*}

\section{Methodology}
\label{sec:method}
In this part, we first introduce the overall architecture of proposed GILA in Section~\ref{subsec:overall_arch}.
Then, we describe its two main components, \textit{i.e.}, the cross-modal global interaction model in Section~\ref{subsec:global_interact} and local alignment approach in Section~\ref{subsec:local_align}.
Finally, we explain the training objective in Section~\ref{subsec:train_obj}.

\subsection{Overall Architecture}
\label{subsec:overall_arch}
As illustrated in Figure~\ref{fig1}(a), the proposed GILA system consists of front-end module, fusion module and recognition module. 
We first introduce a front-end module to pre-process the synchronized audio-video input streams, which employs a linear projection layer for audio front-end and a modified ResNet-18~\cite{shi2022learning} for visual front-end. 
We also concatenate the processed A-V features to build a bottleneck feature ${X}_{B\hspace{-0.02cm}N}$ to collect multimodal information.
Then, we propose a fusion module for audio-visual fusion.
Specifically, we propose a global interaction model and a local alignment approach to capture deep A-V correlations.
The resulted audio, visual and bottleneck features are then concatenated to generate the multimodal feature ${X}_{M\!M}$.
Finally, we introduce a Transformer-based recognition module to encode the multimodal feature and predict the output tokens.
The overall training objective consists of the speech recognition loss $\mathcal{L}_{A\hspace{-0.01cm}S\hspace{-0.01cm}R}$ and the local alignment loss $\mathcal{L}_{L\hspace{-0.01cm}A}$.

\subsection{Cross-Modal Global Interaction (GI)}
\label{subsec:global_interact}
As shown in Figure~\ref{fig1}(b), we propose a cross-modal global interaction model to capture the complementary relationship between A-V modalities. 
Specifically, we first introduce cross-attention to interact audio-visual features to capture inter-modal correspondence.
On top of that, we further propose a novel iterative refinement (IR) module to jointly model the inter- and intra-modal correspondence, aiming to better capture the complementary relationship on modality level.

\vspace{0.1cm}
\noindent\textbf{Cross-Attention} aims to capture the A-V inter-modal correspondence. 
As illustrated in Figure~\ref{fig1}(b), the input audio-visual features of $i$-th GI model layer (\textit{i.e.}, ${X}_{A}^{i-1}$, ${X}_{V}^{i-1}, i\in \{1, 2, 3\}$) are first sent into two separate self-attention modules~\cite{vaswani2017attention} for modeling, which generates two intermediate features, ${F}_{A}^i$ and ${F}_{V}^i$:
\begin{equation}
\label{eq1}
\begin{split}
    {F}_{A}^i &= L\hspace{-0.02cm}N({X}_{A}^{i-1} + M\!H\!A({X}_{A}^{i-1}, {X}_{A}^{i-1}, {X}_{A}^{i-1})),
    \\
    {F}_{V}^i &= L\hspace{-0.02cm}N({X}_{V}^{i-1} + M\!H\!A({X}_{V}^{i-1}, {X}_{V}^{i-1}, {X}_{V}^{i-1})),
\end{split}
\end{equation}
\noindent where ``LN'' denotes layer normalization~\cite{ba2016layer}, ``MHA'' denotes multi-head scaled dot-product attention~\cite{vaswani2017attention}.

Then, we introduce cross-attention to enable audio-visual features to attend to each other for complementation, in order to capture the inter-modal correspondence:
\begin{equation}
\label{eq2}
\begin{split}
    {H}_{A}^i &= L\hspace{-0.02cm}N({F}_{A}^{i} + M\!H\!A({F}_{A}^{i}, {F}_{V}^{i}, {F}_{V}^{i})),
    \\
    {H}_{V}^i &= L\hspace{-0.02cm}N({F}_{V}^{i} + M\!H\!A({F}_{V}^{i}, {F}_{A}^{i}, {F}_{A}^{i})),
\end{split}
\end{equation}

After that, we utilize position-wise feed-forward network (FFN)~\cite{vaswani2017attention} to generate outputs:
\begin{equation}
\label{eq3}
\begin{split}
    {X}_{A}^i &= L\hspace{-0.02cm}N({H}_{A}^{i} + F\!F\!N({H}_{A}^{i}),
    \\
    {X}_{V}^i &= L\hspace{-0.02cm}N({H}_{V}^{i} + F\!F\!N({H}_{V}^{i}),
\end{split}
\end{equation}
\noindent where FFN consists of two linear layers with a ReLU~\cite{glorot2011deep} activation in between.

\begin{figure*}[ht]
\centering
\includegraphics[width=1.0\textwidth]{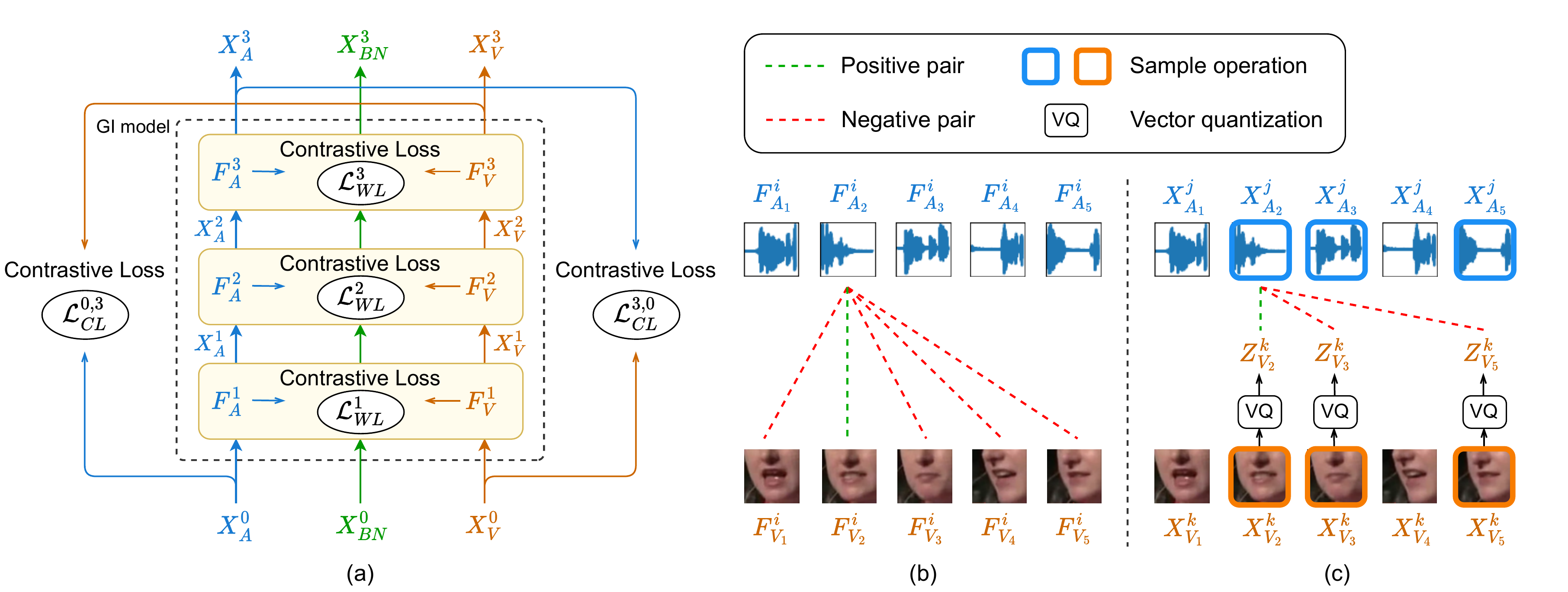}
\vspace{-0.5cm}
\caption{Block diagrams of proposed cross-modal local alignment approach: (a) Overview, (b) Within-Layer (WL) contrastive learning, (c) Cross-Layer (CL) contrastive learning.}\label{fig2}
\vspace{-0.2cm}
\end{figure*}

\vspace{0.1cm}
\noindent\textbf{Iterative Refinement (IR)} aims to jointly model the A-V inter- and intra-modal correspondence, where the bottleneck feature plays a key role.
As shown in Figure~\ref{fig1}(c), the input bottleneck feature ${X}_{B\hspace{-0.02cm}N}^{i-1}$ first attends to the A/V feature from cross-attention (\textit{i.e.}, ${X}_{A}^{i}$, ${X}_{V}^{i}$) respectively, followed by convolution to generate two residual features ${R}_{A}^i$ and ${R}_{V}^i$:
\begin{equation}
\label{eq4}
\begin{split}
    {R}_{A}^i &= \text{Conv}(\text{Attention}({X}_{B\hspace{-0.02cm}N}^{i-1}, {X}_{A}^{i}, {X}_{A}^{i})),
    \\
    {R}_{V}^i &= \text{Conv}(\text{Attention}({X}_{B\hspace{-0.02cm}N}^{i-1}, {X}_{V}^{i}, {X}_{V}^{i})),
\end{split}
\end{equation}
\noindent where ``Conv'' denotes a $1\times1$ convolution layer followed by batch normalization (BN)~\cite{ioffe2015batchnorm} and parametric ReLU (PReLU) activation.

The attention blocks aim to build interactions between individual audio/visual feature and the bottleneck feature that contains multimodal information.
Therefore, the individual A/V modality can not only attend to the other modality, but also attend to itself simultaneously.
As a result, we can jointly model the inter- and intra-modal correspondence, which helps extract the useful information in A/V modality.

Finally, we add the two generated residual features to input bottleneck feature, in order to refine more informative multimodal representations:
\begin{equation}
\label{eq5}
\begin{split}
    {X}_{B\hspace{-0.02cm}N}^i &= L\hspace{-0.02cm}N({X}_{B\hspace{-0.02cm}N}^{i-1} + {R}_{A}^i + {R}_{V}^i),
\end{split}
\end{equation}

With increasing multimodal information in the bottleneck feature, the IR module in next GI model layer can better capture the A-V correspondences by Equation~\ref{eq4}, and so on.
Such refining mechanism enables IR module to effectively model the inter- and intra-modal correspondence.

\subsection{Cross-Modal Local Alignment (LA)}
\label{subsec:local_align}
In order to learn more local details of A-V correlations, we further propose a cross-modal local alignment approach to model the temporal consistency between A-V frames, as presented in Figure~\ref{fig2}.
Specifically, we first introduce within-layer contrastive learning to align the A-V features within same GI model layer.
Based on that, we propose a novel cross-layer contrastive learning method for A-V alignment across different GI model layers, aiming to learn more informative A-V temporal consistency on frame level.

\vspace{0.1cm}
\noindent\textbf{Within-Layer (WL) Contrastive Learning} aims to align the A-V features within same GI model layer.
As illustrated by Figure~\ref{fig2}(a)(b), we select the $i$-th layer's intermediate features ${F}_{A}^i$ and ${F}_{V}^i$ for alignment.
Denote that ${F}_{A}^i = \{{F}_{{A}_t}^i|_{t=1}^{T}\}, {F}_{V}^i = \{{F}_{{V}_t}^i|_{t=1}^{T}\}$, $i\in \{1,2,3\}$, $T$ is number of frames. 
Given each audio frame ${F}_{{A}_t}^i$, the model needs to identify its corresponding visual frame ${F}_{{V}_t}^i$ from the entire visual sequence, and vice versa.
In this sense, the A-V sequences can get well aligned to each other.

\noindent The within-layer contrastive loss is defined as:
\begin{equation}
\label{eq6}
\begin{split}
    \mathcal{L}^{a2v} ({F}_{A}^i, {F}_{V}^i) &= -\sum_{t=1}^{T}\log\frac{\exp(\hspace{0.05cm}\langle{F}_{{A}_t}^i, {F}_{{V}_t}^i\rangle\hspace{0.02cm}/\tau\hspace{0.02cm})}{\sum_{n=1}^{T}\exp(\hspace{0.05cm}\langle{F}_{{A}_t}^i, {F}_{{V}_n}^i\rangle\hspace{0.02cm}/\tau\hspace{0.02cm})},
    \\
    \mathcal{L}^{v2a} ({F}_{V}^i, {F}_{A}^i) &= -\sum_{t=1}^{T}\log\frac{\exp(\hspace{0.05cm}\langle{F}_{{V}_t}^i, {F}_{{A}_t}^i\rangle\hspace{0.02cm}/\tau\hspace{0.02cm})}{\sum_{n=1}^{T}\exp(\hspace{0.05cm}\langle{F}_{{V}_t}^i, {F}_{{A}_n}^i\rangle\hspace{0.02cm}/\tau\hspace{0.02cm})},
    \\
    \mathcal{L}_{W\hspace{-0.02cm}L}^i &= \left[\mathcal{L}^{a2v} ({F}_{A}^i, {F}_{V}^i) + \mathcal{L}^{v2a} ({F}_{V}^i, {F}_{A}^i)\right]/2,
\end{split}
\end{equation}

\noindent where $\langle\hspace{0.04cm}\cdot, \cdot\hspace{0.04cm}\rangle$ denotes cosine similarity, $\tau$ is temperature parameter.
The two alignment directions (\textit{i.e.}, $a2v$, $v2a$) are averaged to obtain the final WL contrastive loss.

\vspace{0.1cm}
\label{subsec:local_align:cl}
\noindent\textbf{Cross-Layer (CL) Contrastive Learning} aims to align the A-V features across different GI model layers.
As presented in Figure~\ref{fig2}(a)(c), we select the $j$-th layer's output audio feature ${X}_{A}^j$ and $k$-th layer's output visual feature ${X}_{V}^k$ for alignment, where $j,k\in\{0,1,2,3\}$, $j\neq k$. 
Particularly, in this work we select $(j, k) \in \{(0,3),(3,0)\}$ to align the input and output A-V features of entire GI model, where more selections are discussed in ablation study (See Section~\ref{subsec:ablation:scenarios}).

Denote that ${X}_{A}^j = \{{X}_{{A}_t}^j|_{t=1}^{T}\}, {X}_{V}^k = \{{X}_{{V}_t}^k|_{t=1}^{T}\}$, where $T$ is number of frames.
First, we randomly sample $T'$ A-V frame pairs from them for alignment, as a dropout to prevent over-fitting.
Therefore, we can write the sampled frames as $\{({X}_{{A}_t}^j, {X}_{{V}_t}^k)|t\in I\}$, where $I \subset \{1, 2, ..., T\}$, $|I| = T'$.

Then, we introduce vector-quantization (VQ)~\cite{baevski2019vq,hu2023wav2code} to discretize the sampled audio-visual frames to a finite set of representations, which results in quantized targets to enable more effective contrastive learning, especially between different-layer features that usually locate in distant domains~\cite{baevski2020wav2vec}:
\begin{equation}
\label{eq7}
\begin{split}
    {Z}_{{A}_t}^j = V\hspace{-0.02cm}Q({X}_{{A}_t}^j),\enspace {Z}_{{V}_t}^k = V\hspace{-0.02cm}Q({X}_{{V}_t}^k),\quad t \in I,
\end{split}
\end{equation}

Finally, we calculate cross-layer contrastive loss to align the audio/visual frames to the quantized visual/audio representations respectively, similar to WL contrastive loss:
\begin{equation}
\label{eq8}
\begin{split}
    \mathcal{L}^{a2v} ({X}_{A}^j, {Z}_{V}^k) &= -\sum_{t\in I}\log\frac{\exp(\hspace{0.05cm}\langle{X}_{{A}_t}^j, {Z}_{{V}_t}^k\rangle\hspace{0.02cm}/\tau\hspace{0.02cm})}{\sum_{n\in I_t}\exp(\hspace{0.05cm}\langle{X}_{{A}_t}^j, {Z}_{{V}_n}^k\rangle\hspace{0.02cm}/\tau\hspace{0.02cm})},\\
    \mathcal{L}^{v2a} ({X}_{V}^k, {Z}_{A}^j) &= -\sum_{t\in I}\log\frac{\exp(\hspace{0.05cm}\langle{X}_{{V}_t}^k, {Z}_{{A}_t}^j\rangle\hspace{0.02cm}/\tau\hspace{0.02cm})}{\sum_{n\in I_t}\exp(\hspace{0.05cm}\langle{X}_{{V}_t}^k, {Z}_{{A}_n}^j\rangle\hspace{0.02cm}/\tau\hspace{0.02cm})},\\
    \mathcal{L}_{C\hspace{-0.02cm}L}^{j,k} &= \left[\mathcal{L}^{a2v} ({X}_{A}^j, {Z}_{V}^k) + \mathcal{L}^{v2a} ({X}_{V}^k, {Z}_{A}^j)\right]/2,
\end{split}
\end{equation}

\noindent where $I_t$ contains the index $t$ and another 100 randomly-selected indexes from $I$, for positive and negative samples respectively~\cite{baevski2020wav2vec}. 
The two alignment directions are averaged to obtain the final CL contrastive loss.

\subsection{Training Objective}
\label{subsec:train_obj}
We first calculate cross-entropy based sequence-to-sequence loss~\cite{watanabe2017hybrid} for speech recognition, as indicated by $\mathcal{L}_{A\hspace{-0.01cm}S\hspace{-0.01cm}R}$ in Figure~\ref{fig1}(a).
Then, we build the local alignment loss $\mathcal{L}_{L\hspace{-0.01cm}A}$ from WL and CL contrastive learning:
\begin{equation}
\label{eq9}
\begin{split}
    \mathcal{L}_{L\hspace{-0.01cm}A} = \sum_{i}^{M}\lambda_{W\hspace{-0.02cm}L}^i \cdot \mathcal{L}_{W\hspace{-0.02cm}L}^i + \sum_{(j,k)}^{N}\lambda_{C\hspace{-0.02cm}L}^{j,k} \cdot \mathcal{L}_{C\hspace{-0.02cm}L}^{j,k}
\end{split}
\end{equation}

\noindent where $M=\{1,2,3\}$, $N=\{(0,3),(3,0)\}$, $\lambda_{W\hspace{-0.02cm}L}^i$ and $\lambda_{C\hspace{-0.02cm}L}^{j,k}$ are weighting parameters for different training objectives.

We combine them to form the final training objective and train the entire GILA system in an end-to-end manner:
\begin{equation}
\label{eq9}
\begin{split}
    \mathcal{L}_{G\hspace{-0.02cm}I\hspace{-0.02cm}L\hspace{-0.02cm}A} = \mathcal{L}_{A\hspace{-0.01cm}S\hspace{-0.01cm}R} + \mathcal{L}_{L\hspace{-0.01cm}A}
\end{split}
\end{equation}

\section{Experiments}
\label{sec:exp}

\subsection{Experimental Setup}
\label{subsec:setup}
\noindent\textbf{Datasets.} We conduct experiments on two large-scale publicly available datasets, LRS3~\cite{afouras2018lrs3} and LRS2~\cite{chung2017lip}. 
LRS3 dataset collects 433 hours of transcribed English videos from TED and TEDx talks. 
LRS2 dataset contains 224 hours of video speech from BBC programs.
More details are in Appendix~\ref{subsec:datasets}.
\vspace{0.05cm}

\label{subsubsec:baselines}
\noindent\textbf{Baselines.} We employ AV-HuBERT\footnote{\url{https://github.com/facebookresearch/av_hubert}}~\cite{shi2022learning} as our baseline, but for fair comparison we discard the pre-training stage. 
To evaluate our GILA, we select some popular AVSR methods for comparison: TM-seq2seq, TM-CTC, Hyb-RNN, EG-seq2seq, RNN-T, LF-MMI TDNN, Hyb-Conformer, MoCo+wav2vec, AV-HuBERT (LARGE), u-HuBERT (LARGE), which are introduced in Section~\ref{sec:related_work}.
\vspace{0.05cm}

\label{subsubsec:implementation_details}
\noindent\textbf{Implementation Details.} For model configurations, our baseline follows AV-HuBERT LARGE~\cite{shi2022learning} with 24 Transformer encoder layers and 9 decoder layers.
For fair comparison, we build the GILA with 3 GI model layers, 12 Transformer encoder layers and 9 decoder layers. 
All other model configurations are same as AV-HuBERT LARGE.
The number of parameters in our baseline and GILA are 476M and 465M respectively.
We also use Conformer as our backbone, with the convolution kernel size of 31.



\begin{table}[h]
\centering
\resizebox{0.98\columnwidth}{!}{
\begin{tabular}{llcccc}
\toprule
\multicolumn{2}{c}{\multirow{2}{*}{Method}} & \multirow{2}{*}{Backbone} & \multirow{2}{*}{LM} & \multicolumn{2}{c}{WER(\%)} \\
\multicolumn{2}{c}{} & & & Clean & Noisy \\
\midrule
\multicolumn{2}{c}{TM-seq2seq~\shortcite{afouras2018deep}} & Transformer & \cmark & 7.2 & - \\
\multicolumn{2}{c}{EG-seq2seq~\shortcite{xu2020discriminative}} & RNN & - & 6.8 & - \\
\multicolumn{2}{c}{RNN-T~\shortcite{makino2019recurrent}} & RNN & - & 4.5 & - \\
\multicolumn{2}{c}{Hyb-Conformer~\shortcite{ma2021end}} & Conformer & \cmark & 2.3* & - \\
\multicolumn{2}{c}{AV-HuBERT~\shortcite{shi2022learning}} & Transformer & - & \hspace{0.175cm}1.4** & \hspace{0.175cm}5.8**\\
\multicolumn{2}{c}{u-HuBERT~\shortcite{hsu2022u}} & Transformer & - & \hspace{0.175cm}1.2** & -\\
\midrule
\multirow{8}{*}{GILA (ours)} & Baseline & \multirow{4}{*}{Transformer} & \multirow{4}{*}{-} & 3.75 & 17.22 \\
& \quad+ GI & & & 3.29 & 15.06 \\
& \quad\quad+ LA & & & 2.88 & 13.35 \\
& \quad\quad\quad+ DA & & & 2.61 & 11.14\\
\cline{2-6}
& Baseline & \multirow{4}{*}{Conformer} & \multirow{4}{*}{-} & 2.64 & 11.89 \\
& \quad+ GI & & & 2.31 & 10.34 \\
& \quad\quad+ LA & & & 2.04 & 8.97 \\
& \quad\quad\quad+ DA & & & \textbf{1.96} & \textbf{7.03} \\
\bottomrule
\end{tabular}
}
\caption{WER (\%) of GILA and prior works on LRS3 banchmark. 
``GI'' denotes global interaction model, ``LA'' denotes local alignment approach, ``DA'' denotes data augmentation.
``LM'' denotes language model rescoring.
* denotes using hybrid seq2seq/CTC loss for training, external LM rescoring for inference and extra data to pre-train the audio/visual front-ends.
** denotes using self-supervised pre-training with extra unlabeled data ($>$ 1,700 hours).} 
\label{table1}
\end{table}

\begin{table}[h]
\centering
\resizebox{0.98\columnwidth}{!}{
\begin{tabular}{llcccc}
\toprule
\multicolumn{2}{c}{\multirow{2}{*}{Method}} & \multirow{2}{*}{Backbone} &
\multirow{2}{*}{LM} &\multicolumn{2}{c}{WER(\%)} \\
\multicolumn{2}{c}{} & & & Clean & Noisy \\
\midrule
\multicolumn{2}{c}{TM-seq2seq~\shortcite{afouras2018deep}} & Transformer & \cmark & 8.5 & - \\
\multicolumn{2}{c}{TM-CTC~\shortcite{afouras2018deep}} & Transformer & \cmark & 8.2 & - \\
\multicolumn{2}{c}{Hyb-RNN~\shortcite{petridis2018audio}} & RNN & \cmark & 7.0 & - \\
\multicolumn{2}{c}{LF-MMI TDNN~\shortcite{yu2020audio}} & TDNN & \cmark & 5.9 & - \\
\multicolumn{2}{c}{Hyb-Conformer~\shortcite{ma2021end}} & Conformer & \cmark & 3.7* & - \\
\multicolumn{2}{c}{MoCo+wav2vec~\shortcite{pan2022leveraging}} & Transformer & - & \hspace{0.175cm}2.6** & - \\
\midrule
\multirow{8}{*}{GILA (ours)} & Baseline & \multirow{4}{*}{Transformer} & \multirow{4}{*}{-} & 5.79 & 25.52 \\
& \quad+ GI & & & 4.98 & 21.91 \\
& \quad\quad+ LA & & & 4.31 & 18.84 \\
& \quad\quad\quad+ DA & & & 4.02 & 15.70\\
\cline{2-6}
& Baseline & \multirow{4}{*}{Conformer} & \multirow{4}{*}{-} & 4.09 & 17.83 \\
& \quad+ GI & & & 3.54 & 15.41 \\
& \quad\quad+ LA & & & 3.17 & 13.75 \\
& \quad\quad\quad+ DA & & & \textbf{3.10} & \textbf{11.24} \\
\bottomrule
\end{tabular}
}
\caption{WER (\%) of our GILA and prior works on the LRS2 benchmark. 
* denotes the same as that in Table~\ref{table1}.
** denotes using self-supervised pre-trained audio/visual front-ends.}
\label{table2}
\end{table}

The system inputs are log filterbank features for audio stream and lip regions-of-interest (ROIs) for video stream.
To sample A-V frame pairs in CL contrastive learning, we first sample starting indexes from $(X_A^0, X_V^3)$ with probability of 0.4 and from $(X_A^3, X_V^0)$ with 0.45 respectively, and then cut out 10 consecutive frames after each sampled index.
To calculate contrastive loss, we use the same VQ module in wav2vec2.0~\cite{baevski2020wav2vec}, and set the temperature parameter $\tau$ to 0.1.
We further use data augmentation to improve noise robustness, where we add MUSAN noise~\cite{snyder2015musan} following prior work~\cite{shi2022robust}, and report WER results on both clean and noisy test sets.
The weighting parameters $\lambda_{W\hspace{-0.02cm}L}^i(i\hspace{-0.03cm}\in\hspace{-0.03cm}\{1,2,3\})/\lambda_{C\hspace{-0.02cm}L}^{0,3}/\lambda_{C\hspace{-0.02cm}L}^{3,0}$ are set to 0.001/0.08/0.01 respectively.
All hyper-parameters are tuned on validation set.
Our training follows the finetuning configurations in~\cite{shi2022learning} and takes $\sim$ 1.3 days on 4 V100-32GB GPUs, which is much more efficient than AV-HuBERT pre-training ($\sim$ 15.6 days on 64 V100-GPUs).
More details of baselines, data augmentation, model and training configurations are presented in Appendix~\ref{sec:exp_details}.

\subsection{Main Results}
\label{subsec:main_results}
\noindent\textbf{Results on LRS3.} Table~\ref{table1} compares the performance of our proposed GILA with existing methods on LRS3 benchmark.
Under clean test set, our best model outperforms the supervised learning SOTA by 14.8\% relatively (2.3\%$\rightarrow$1.96\%), while without the CTC training loss, external LM rescoring and extra A/V front-end pre-training that their method uses.
Moreover, the proposed GILA has also achieved significant WER improvements over our baseline (3.75\%$\rightarrow$2.61\%, 2.64\%$\rightarrow$1.96\%).
Specifically, its two main components, \textit{i.e.}, GI model and LA method, both contribute a lot to the improvements, and the data augmentation also yields better results.
We can also observe similar improvements on noisy test set.
In addition, the Conformer backbone significantly outperforms Transformer (2.61\%$\rightarrow$1.96\%).

\vspace{0.05cm}
\noindent\textbf{Results on LRS2.} Table~\ref{table2} compares the performance of our GILA with existing AVSR methods on LRS2 benchmark.
Under clean test set, our best model achieves 16.2\% relative WER improvement over the supervised learning SOTA (3.7\%$\rightarrow$3.10\%).
Moreover, the GILA has also achieved significant improvements over our baseline (5.79\%$\rightarrow$4.02\%, 4.09\%$\rightarrow$3.10\%), where the GI model, LA method and data augmentation all yield positive contributions.

Therefore, our GILA has achieved new supervised learning SOTA on both LRS3 and LRS2 benchmarks, with up to 16.2\% relative WER improvement over the best baseline. 
It also moves closer to the self-supervised learning SOTA (1.96\% vs. 1.2\%, 3.10\% vs. 2.6\%) while costs no unlabeled data and much less computing resources (See Section~\ref{subsubsec:implementation_details}).

\begin{table}[t]
\centering
\resizebox{0.92\columnwidth}{!}{
\begin{tabular}{lccc}
\toprule
\multirow{2}{*}{Method} & \multirow{2}{*}{Backbone} & \multicolumn{2}{c}{WER(\%)} \\
& & Clean & Noisy \\
\midrule
Baseline & \multirow{4}{*}{Transformer-LARGE} & 3.75 & 17.22 \\
\quad+ cross-attention & & 3.50 & 15.90 \\
\quad+ IR module & & 3.61 & 16.48 \\
\quad+ both (GI) & & 3.29 & 15.06 \\
\midrule
Baseline & \multirow{4}{*}{Conformer-LARGE} & 2.64 & 11.89 \\
\quad+ cross-attention & & 2.45 & 10.94 \\
\quad+ IR module & & 2.53 & 11.41 \\
\quad+ both (GI) & & \textbf{2.31} & \textbf{10.34} \\
\bottomrule
\end{tabular}
}
\caption{Effect of global interaction (GI) model and its two sub-modules on LRS3 benchmark.
``+ cross-attention'' denotes using cross-attention module separately, ``+ IR module'' denotes using iterative refinement module separately, where the self-attention and FFN modules in GI model are always maintained.}
\label{table3}
\end{table}

\begin{figure}[t]
\centering
\includegraphics[width=1.0\columnwidth]{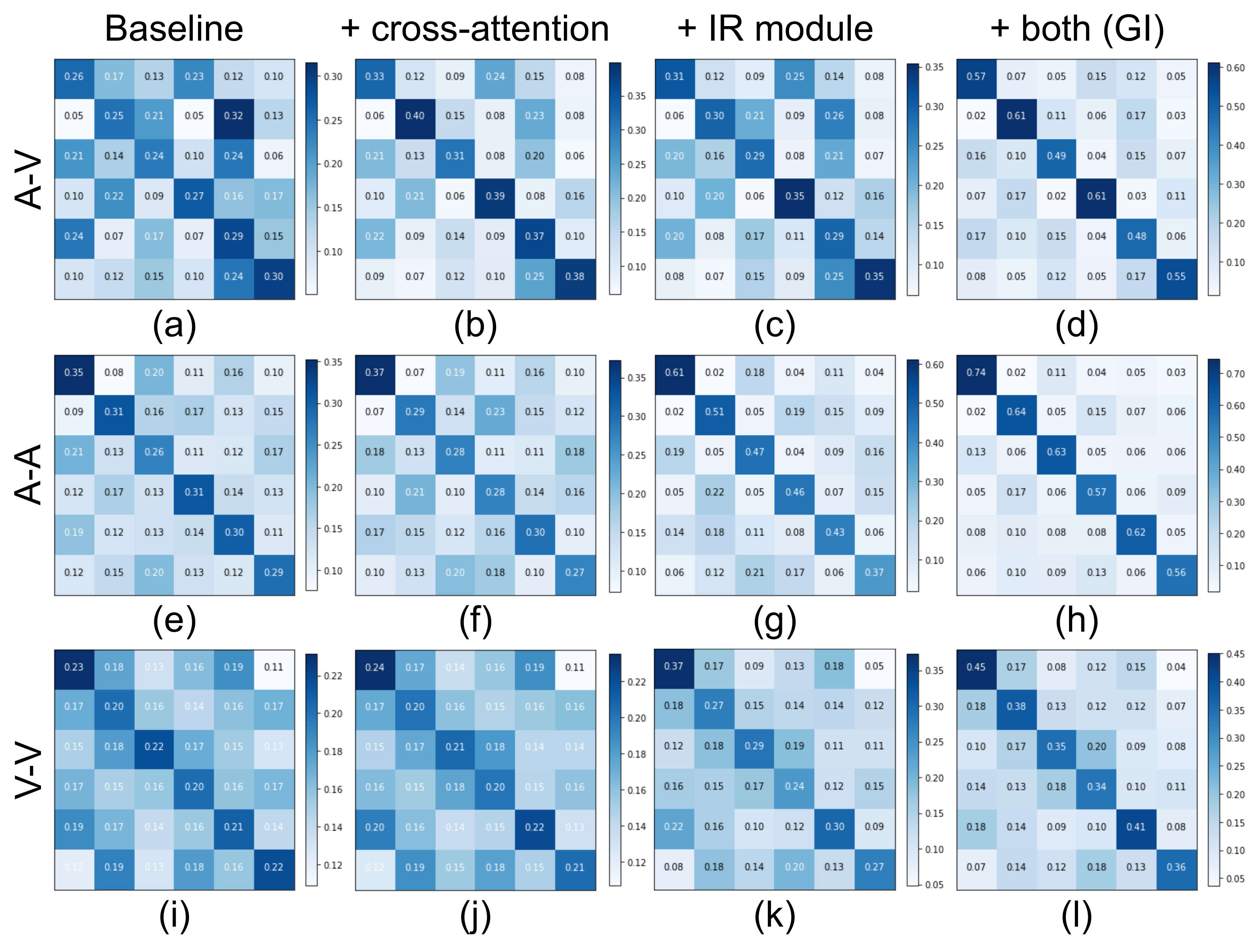}
\vspace{-0.45cm}
\caption{Cosine similarity matrix (after softmax) between audio-visual (row 1), audio-audio (row 2) and visual-visual (row 3) sequence embeddings in GI model.
The column 1-4 denotes baseline, baseline + cross-attention, baseline + IR module, baseline + both (GI), respectively.
In row 1, horizontal axis denotes visual sequences in a batch and vertical axis denotes the audio sequences, which are selected from LRS3 test set.
Sequence embedding is obtained by temporal pooling on the output audio/visual sequences, \textit{i.e.}, ${X}_{A}^3, {X}_{V}^3$.}\label{fig3}
\vspace{-0.05cm}
\end{figure}

\subsection{Ablation Study}
\label{subsec:ablation}
\noindent\textbf{Effect of Global Interaction Model.}
Table~\ref{table3} summarizes the effect of proposed GI model and its two sub-modules, \textit{i.e.}, cross-attention and IR modules.
We first observe that using cross-attention to capture inter-modal correspondence can improve the WER results (3.75\%$\rightarrow$3.50\%, 2.64\%$\rightarrow$2.45\%). 
Further improvements can be achieved by adding IR module to jointly model the inter- and intra-modal correspondence (3.50\%$\rightarrow$3.29\%, 2.45\%$\rightarrow$2.31\%), where using it separately can also improve.
Similar improvements can be observed on the noisy test set.
Therefore, these results verify the effectiveness of our proposed GI model.

\vspace{0.05cm}
\noindent\textbf{Visualizations of Inter- and Intra-Modal Correspondence.} 
Figure~\ref{fig3} visualizes the captured inter- and intra-modal correspondence by our GI model, using similarity matrixes where the diagonal elements denote cosine similarity between true A-V, A-A or V-V pairs.
We first observe chaotic mappings between A-V embeddings in baseline from Figure~\ref{fig3}(a).
After introducing cross-attention to interact A-V features, we can capture some inter-modal correspondence between true A-V pairs, \textit{i.e.}, (b) vs. (a).
However, it fails to capture the A/V intra-modal correspondence, \textit{i.e.}, (f) vs. (e), (j) vs. (i).
Thus, we further propose an iterative refinement module to jointly model the inter- and intra-modal correspondence, which improves significantly as indicated by the clearer diagonals in column 4.
As a result, our GI model can effectively capture both inter- and intra-modal correspondence.

We further investigate the relationship between these two correspondences.
When compared to baseline, using cross-attention can learn better inter-modal correspondence, \textit{i.e.}, (b) vs. (a), while using it on top of IR module achieves significantly more improvements, \textit{i.e.}, (d) vs. (c).
Similar phenomenon can be observed on WER results in Table~\ref{table3}.
It indicates that the proposed IR could be beneficial to cross-attention, where its captured intra-modal correspondence could help to model the inter-modal correspondence, thus results in better A-V complementary relationship.

\noindent\textbf{Effect of Local Alignment Approach.}
Table~\ref{table4} summarizes the effect of proposed LA method and its two components, \textit{i.e.}, within-layer and cross-layer contrastive learning.
We first introduce WL contrastive learning for audio-visual alignment within same GI model layer, which can improve the WER performance (3.29\%$\rightarrow$3.03\%, 2.31\%$\rightarrow$2.13\%).
Further improvements can be achieved by adding CL contrastive learning to align the A-V features across different layers (3.03\%$\rightarrow$2.88\%, 2.13\%$\rightarrow$2.04\%), where using it separately can also improve.
Similar improvements can be observed on noisy test set.
Therefore, these results validate the effectiveness of our proposed LA method.

\begin{table}[t]
\centering
\resizebox{1.0\columnwidth}{!}{
\begin{tabular}{lccc}
\toprule
\multirow{2}{*}{Method} & \multirow{2}{*}{Backbone} & \multicolumn{2}{c}{WER(\%)} \\
& & Clean & Noisy \\
\midrule
GI model & \multirow{4}{*}{Transformer-LARGE} & 3.29 & 15.06 \\
\quad+ WL contrastive learning & & 3.03 & 13.92 \\
\quad+ CL contrastive learning & & 3.11 & 14.36 \\
\quad+ both (LA) & & 2.88 & 13.35 \\
\midrule
GI model & \multirow{4}{*}{Conformer-LARGE} & 2.31 & 10.34 \\
\quad+ WL contrastive learning & & 2.13 & 9.53 \\
\quad+ CL contrastive learning & & 2.18 & 9.70 \\
\quad+ both (LA) & & \textbf{2.04} & \textbf{8.97} \\
\bottomrule
\end{tabular}
}
\caption{Effect of local alignment (LA) approach and its two components on LRS3 benchmark.}
\label{table4}
\end{table}

\begin{table}[t]
\centering
\resizebox{0.58\columnwidth}{!}{
\begin{tabular}{c|cccc}
\toprule
WER(\%) & $X_V^0$ & $X_V^1$ & $X_V^2$ & $X_V^3$ \\
\midrule
$X_A^0$ & - & 2.12 & 2.11 & \textbf{2.07} \vspace{0.15cm}\\
$X_A^1$ & 2.12 & - & 2.12 & 2.09 \vspace{0.15cm}\\
$X_A^2$ & 2.09 & 2.10 & - & 2.11 \vspace{0.15cm}\\
$X_A^3$ & \textbf{2.06} & 2.08 & 2.10 & - \\
\bottomrule
\end{tabular}
}
\caption{Effect of cross-layer contrastive learning. 
We select different A-V feature pairs $({X}_{A}^j, {X}_{V}^k)$ for cross-layer alignment. 
The baseline we use in this study is GI model with WL contrastive learning (2.13\% WER in Table~\ref{table4}).}
\vspace{-0.3cm}
\label{table5}
\end{table}

\vspace{0.05cm}
\label{subsec:ablation:scenarios}
\noindent\textbf{Effect of Cross-Layer Contrastive Learning.}
Table~\ref{table5} further analyzes the effect of cross-layer contrastive learning, where we report WER results of alignment between different A-V feature pairs $({X}_{A}^j, {X}_{V}^k)$.
We observe that the more layers our A-V alignment across (\textit{i.e.}, larger $|j-k|$), the better performance we can achieve, where the best two results (2.07\%, 2.06\%) are achieved by aligning the input and output A-V features of entire GI model.
After combining them, we can achieve even better WER result, as indicated in Table~\ref{table4} (2.04\%).
The reason could be that, the higher-layer features contain semantic representations of larger granularity, or larger receptive field.
Therefore, the A-V alignment across more layers also means across larger granularity gap, which could learn richer cross-modal contextual information and results in more informative A-V temporal consistency.

\begin{figure}[t]
\centering
\includegraphics[width=1.0\columnwidth]{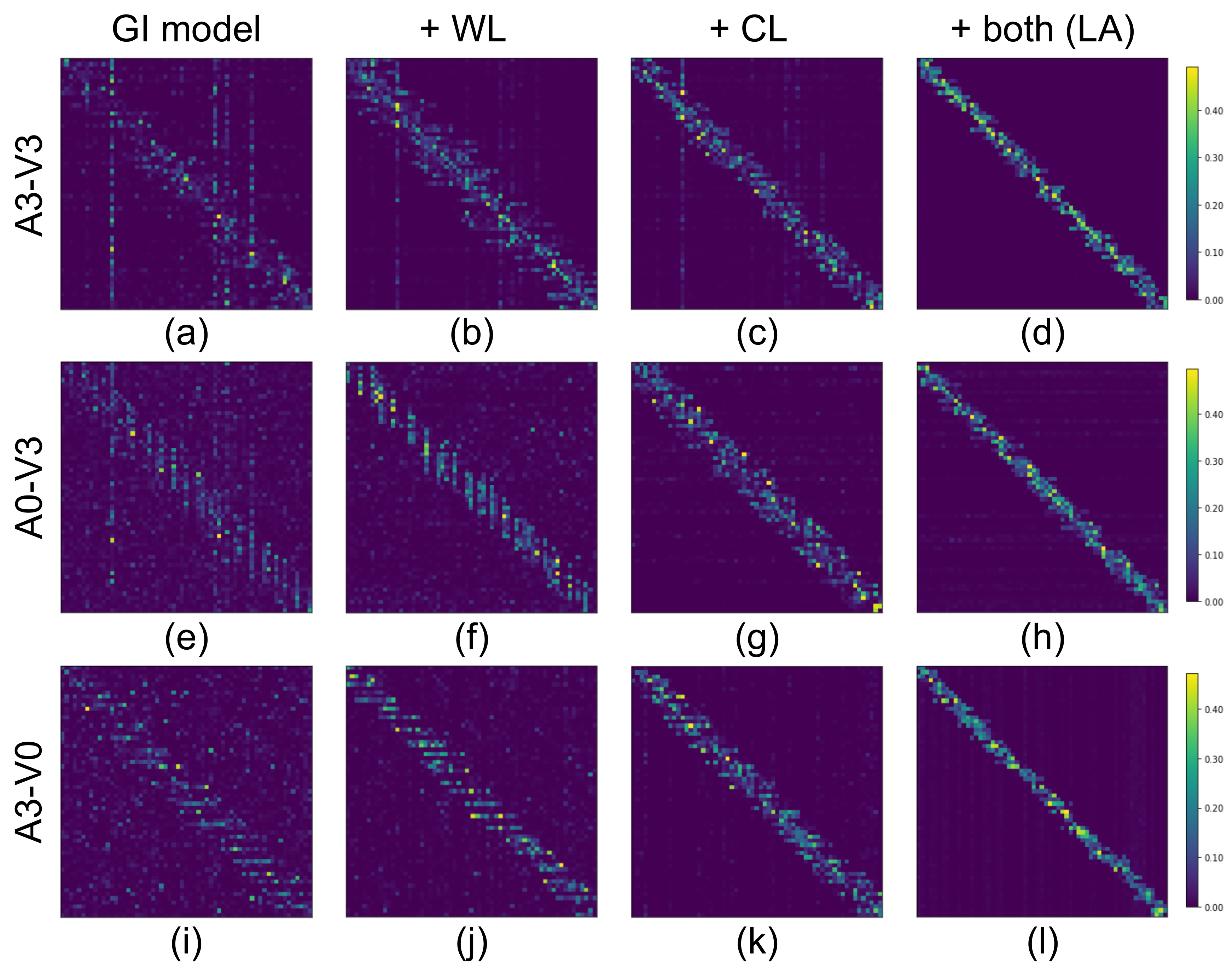}
\vspace{-0.44cm}
\caption{Attention weight map between different audio-visual sequences with LA method: row 1: $({X}_{A}^3, {X}_{V}^3)$, row 2: $({X}_{A}^0, {X}_{V}^3)$, row 3: $({X}_{A}^3, {X}_{V}^0)$.
The column 1-4 denotes GI model, GI + WL contrastive learning, GI + CL contrastive learning, GI + both (LA), respectively.
The x-axis denotes visual frames in an utterance and y-axis denotes the audio frames in utterance, which is selected from LRS3 test set.}\label{fig4}
\vspace{-0.26cm}
\end{figure}

\vspace{0.05cm}
\noindent\textbf{Visualizations of Audio-Visual Temporal Consistency.} 
Figure~\ref{fig4} visualizes the A-V temporal consistency modeled by within-layer and cross-layer contrastive learning, using attention map where the diagonal elements indicate the attention weights between corresponding A-V frames.
We first observe misalignment between A-V sequences in GI model, such as the one-to-many lip-audio mappings shown in Figure~\ref{fig4}(a).
Our proposed WL contrastive learning can help model the temporal consistency between A-V sequences, as indicated by the clearer diagonal in (b).
Similar improvements can be observed on cross-layer temporal consistency, \textit{i.e.}, (f)/(j) vs. (e)/(i), while we also observe some vertical and horizontal stripes near the diagonal, which indicate the granularity gap between different-layer features.

Then in the proposed CL contrastive learning that consists of two alignment directions (See Equation~\ref{eq8}), the low-layer features first learn rich A-V contextual correlations from the high-layer features that with large receptive field, which alleviates the granularity gap between them, \textit{i.e.}, (g)/(k) vs. (e)/(i), (h)/(l) vs. (f)/(j).
Meanwhile, the high-layer features can learn clearer A-V contextual mappings by aligned to the low-layer features that with small granularity, as indicated by the brighter diagonals in Figure~\ref{fig4} (column 3 vs. column 1, column 4 vs. column 2).
As a result, the proposed cross-layer alignment can capture rich cross-modal contextual information to learn better A-V temporal consistency.

\section{Conclusion}
\label{sec:conclusion}
In this paper, we propose a cross-modal global interaction and local alignment (GILA) approach for audio-visual speech recognition, in order to capture the deep audio-visual correlations from both global and local perspectives.
In particular, we first propose a global interaction model to capture the A-V complementary relationship on modality level.
Furthermore, we design a cross-modal local alignment approach to model the A-V temporal consistency on frame level.
Such a holistic view of cross-modal correlations enable better multimodal representations for AVSR.
Experimental results on two public benchmarks demonstrate that our approach has achieved the state-of-the-art in supervised learning methods.

\section*{Acknowledgments}
This research is supported by ST Engineering Mission Software \& Services Pte. Ltd under collaboration programme (Research Collaboration No.: REQ0149132).
The computational work for this article was partially performed on resources of the National Supercomputing Centre, Singapore (\url{https://www.nscc.sg}).


\bibliographystyle{named}

\appendix

\section{Experimental Details}
\label{sec:exp_details}

\subsection{Datasets}
\label{subsec:datasets}
\noindent\textbf{LRS3}\footnote{\url{https://www.robots.ox.ac.uk/~vgg/data/lip_reading/lrs3.html}}~\cite{afouras2018lrs3} is currently the largest public sentence-level lip reading dataset, which contains over 400 hours of English video extracted from TED and TEDx talks on YouTube.
The training data is divided into two parts: pretrain (403 hours) and trainval (30 hours), and both of them are transcribed at sentence level.
The pretrain part differs from trainval in that the duration of its video clips are at a much wider range. 
Since there is no official development set provided, we randomly select 1,200 samples from trainval as validation set ($\sim$ 1 hour) for early stopping and hyper-parameter tuning.
In addition, it provides a standard test set (0.9 hours) for evaluation.

\vspace{0.05cm}
\noindent\textbf{LRS2}\footnote{\url{https://www.robots.ox.ac.uk/~vgg/data/lip_reading/lrs2.html}}~\cite{chung2017lip} is a large-scale publicly available labeled audio-visual (A-V) datasets, which consists of 224 hours of video clips from BBC programs.
The training data is divided into three parts: pretrain (195 hours), train (28 hours) and val (0.6 hours), which are all transcribed at sentence level.
An official test set (0.5 hours) is provided for evaluation use.
The dataset is very challenging as there are large variations in head pose, lighting conditions and genres.

\subsection{Data Preprocessing}
\label{subsec:preprocessing}
The data preprocessing for above two datasets follows the LRS3 preprocessing steps in prior work\footnote{\url{https://github.com/facebookresearch/av_hubert/tree/main/avhubert/preparation}}~\cite{shi2022learning}.
For the audio stream, we extract the 26-dimensional log filter-bank feature at a stride of 10 ms from input raw waveform.
For the video clips, we detect the 68 facial keypoints using dlib toolkit~\cite{king2009dlib} and align the image frame to a reference face frame via affine transformation.
Then, we convert the image frame to gray-scale and crop a 96$\times$96 region-of-interest (ROI) centered on the detected mouth.
During training, we randomly crop a 88$\times$88 region from the whole ROI and flip it horizontally with a probability of 0.5. 
At inference time, the 88$\times$88 ROI is center cropped without horizontal flipping. 
To synchronize these two modalities, we stack each 4 neighboring acoustic frames to match the image frames that are sampled at 25Hz.

\begin{table}[t]
\centering
\resizebox{0.8\columnwidth}{!}{
\begin{tabular}{ccc}
\toprule
Method & Backbone & \# Params.(M) \\
\midrule
\multirow{2}{*}{Baseline} & Transformer-LARGE & 476 \vspace{0.05cm} \\
& Conformer-LARGE & 587 \vspace{0.01mm} \\
\midrule
\multirow{2}{*}{GILA (ours)} & Transformer-LARGE & 465 \vspace{0.05cm} \\
& Conformer-LARGE & 529 \vspace{0.01mm} \\
\bottomrule
\end{tabular}
}
\vspace{-0.05cm}
\caption{Number of parameters in different configurations.}
\vspace{-0.3cm}
\label{table6}
\end{table}

\subsection{Model Configurations}
\label{subsec:model_config}
\noindent\textbf{Front-end.} We use one linear projection layer followed by layer normalization~\cite{ba2016layer} as the audio front-end.
For video front-end, we adopt the modified ResNet-18 from prior work~\cite{shi2022learning}, where the first convolutional layer is replaced by a 3D convolutional layer with kernel size of 5$\times$7$\times$7. 
The visual feature is squeezed into an 1D tensor by spatial average pooling in the end.

\vspace{0.05cm}
\noindent\textbf{GILA Architecture.}
Our baseline is borrowed from AV-HuBERT model~\cite{shi2022learning}, which contains 24 Transformer encoder layers and 9 Transformer decoder layers.
To maintain similar model size, our proposed GILA contains 3 GI model layers, 12 Transformer encoder layers and 9 Transformer decoder layers.
The embedding dimension/feed-forward dimension/attention heads in each Transformer layer are set to 1024/4096/16 respectively, where we use a dropout for each self-attention block at rate of 0.1.
In addition to Transformer, we also employ Conformer~\cite{gulati2020conformer} as our backbone, where we set the depth-wise convolution kernel size to 31.
The Conformer-based GI model consists of FFN, self-attention, cross-attention, convolution module and FFN in sequential.
To save model size and prevent over-fitting in Conformer backbone, we set the inner dimension of convolution module to 128 and the feed-forward dimension to 3072.
Number of parameters in all configurations are presented in Table~\ref{table6}.

\begin{figure}[t]
\centering
\includegraphics[width=0.64\columnwidth]{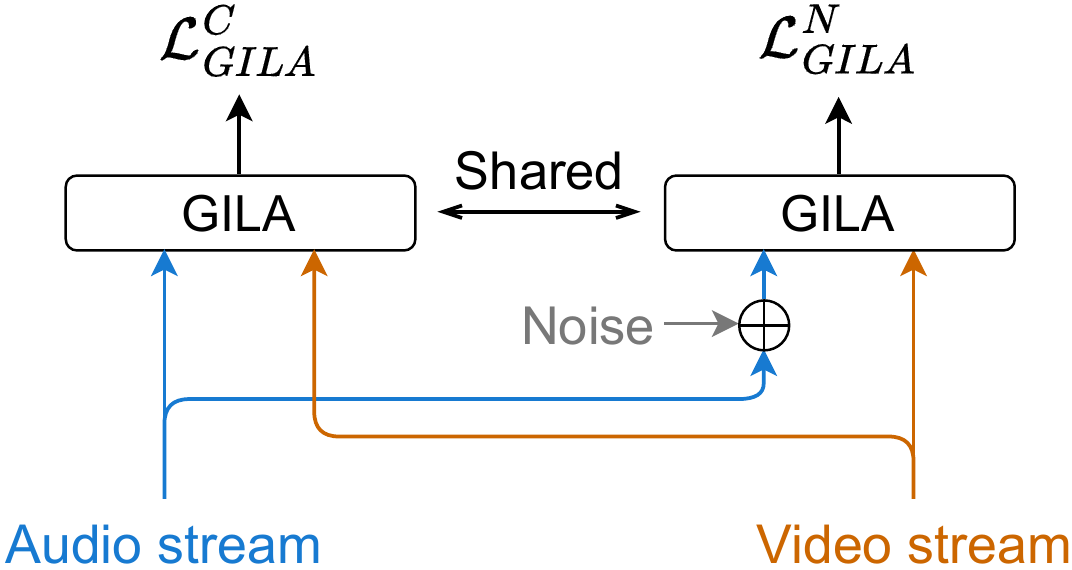}
\vspace{-0.1cm}
\caption{Block diagram of data augmentation.}\label{fig5}
\vspace{-0.3cm}
\end{figure}

\subsection{Noise and Data Augmentation}
\label{subsec:data_aug}
We use many noise categories for noise and data augmentation.
We first select the noise categories of ``\texttt{natural}'', ``\texttt{music}'' and ``\texttt{babble}'' from MUSAN noise dataset~\cite{snyder2015musan}, and then extract some overlapping ``\texttt{speech}'' noise samples from LRS3 dataset.
All categories are divided into training, validation and test partitions, following the prior work~\cite{shi2022robust}.

We define two augmentation techniques in this work, \textit{i.e.}, noise augmentation and data augmentation.
For noise augmentation, we randomly select one noise category and sample a noise clip from its training partition. 
Then, we randomly mix the sampled noise with input clean audio, at 0dB SNR with a probability of 0.25.
Based on that, for data augmentation we feed both the clean and noise-augmented audios into GILA for system training, where they are paired with same corresponding video input, as shown in Figure~\ref{fig5}.
These two data flows (\textit{i.e.}, clean and noisy) share the GILA model parameters, which results in two training objectives from Equation 10 in the main paper, $\mathcal{L}_{G\hspace{-0.02cm}I\hspace{-0.02cm}L\hspace{-0.02cm}A}^C$ and $\mathcal{L}_{G\hspace{-0.02cm}I\hspace{-0.02cm}L\hspace{-0.02cm}A}^N$.
Finally, these two losses are weight summed for multi-task learning:
\begin{equation}
\label{eq1}
\begin{split}
    \mathcal{L}_{final} &= \lambda_C \cdot \mathcal{L}_{G\hspace{-0.02cm}I\hspace{-0.02cm}L\hspace{-0.02cm}A}^C + (1-\lambda_C) \cdot \mathcal{L}_{G\hspace{-0.02cm}I\hspace{-0.02cm}L\hspace{-0.02cm}A}^N
\end{split}
\end{equation}
\noindent where the weighting parameter $\lambda_{C}$ is set to 0.6. The entire system is trained in an end-to-end manner.
We use noise augmentation technique everywhere without specified, otherwise we use the data augmentation technique.

At inference time, we evaluate our model on clean and noisy test sets respectively.
Specifically, the model performance on each noise type is evaluated separately, where the testing noise clips are added at five different SNR levels: $\{-10, -5, 0, 5, 10\}dB$.
At last, the testing results on different noise types and SNR levels will be averaged to obtain the final noisy WER result.

\subsection{Training and Inference}
\label{subsec:train_infer_config}
\noindent\textbf{Training.} We follow the sequence-to-sequence (S2S) finetuning configurations of AV-HuBERT~\cite{shi2022robust} to train our systems.
We use Transformer decoder to decode the encoded features into unigram-based subword units~\cite{kudo2018subword}, where the vocabulary size is set to 1000.
The entire system is trained for 60K steps using Adam optimizer~\cite{kingma2014adam}, where the learning rate is warmed up to a peak of 0.001 for the first 20K updates and then linearly decayed.
The training process takes $\sim$ 1.3 days on 4 NVIDIA-V100-32GB GPUs, which is much more efficient than AV-HuBERT pre-training ($\sim$ 15.6 days on 64 V100-GPUs).

\vspace{0.05cm}
\noindent\textbf{Inference.} No language model is used during inference.
We employ beam search for decoding, where the beam width and length penalty are set to 50 and 1 respectively.
All hyper-parameters in our systems are tuned on validation set.

\subsection{Baselines}
\label{subsec:baselines}
In this section, we describe the baselines for comparison.
\begin{itemize}
\item \textbf{TM-seq2seq}~\cite{afouras2018deep}: TM-seq2seq proposes a Transformer-based~\cite{vaswani2017attention} AVSR system to model the A-V features separately and then attentively fuse them for decoding, and uses sequence-to-sequence loss~\cite{watanabe2017hybrid} as training criterion.
\item \textbf{TM-CTC}~\cite{afouras2018deep}: TM-CTC shares the same architecture with TM-seq2seq, but uses CTC loss~\cite{graves2006connectionist} as training criterion.
\item \textbf{Hyb-RNN}~\cite{petridis2018audio}: Hyb-RNN proposes a RNN-based AVSR model with hybrid seq2seq/CTC loss~\cite{watanabe2017hybrid}, where the A-V features are encoded separately and then concatenated for decoding.
\item \textbf{RNN-T}~\cite{makino2019recurrent}: RNN-T adopts the popular recurrent neural network transducer~\cite{graves2012sequence,liu2021ustc} for AVSR task, where the audio and visual features are concatenated before fed into the encoder.
\item \textbf{EG-seq2seq}~\cite{xu2020discriminative}: EG-seq2seq builds a joint audio enhancement~\cite{zhu2022noise,zhu2023joint,chen2023metric,chen2023unsupervised,hu2023gradient} and multimodal speech recognition system based on the element-wise attention gated recurrent unit (EleAtt-GRU)~\cite{zhang2019eleatt}, where the A-V features are concatenated before decoding.
\item \textbf{LF-MMI TDNN}~\cite{yu2020audio}: LF-MMI TDNN proposes a joint audio-visual speech separation and recognition system~\cite{hu2023unifying} based on time-delay neural network (TDNN), where the A-V features are concatenated before fed into the recognition network.
\item \textbf{Hyb-Conformer}~\cite{ma2021end}: Hyb-Conformer proposes a Conformer-based~\cite{gulati2020conformer} AVSR system with hybrid seq2seq/CTC loss, where the A-V input streams are first encoded separately and then concatenated for decoding.
\item \textbf{MoCo+wav2vec}~\cite{pan2022leveraging}: MoCo+wav2vec employs self-supervised pre-trained audio and visual front-ends, \textit{i.e.}, wav2vec 2.0~\cite{baevski2020wav2vec} and MoCo v2~\cite{chen2020improved}, to generate better audio-visual features for fusion and decoding.
\item \textbf{AV-HuBERT}~\cite{shi2022learning,shi2022robust}: AV-HuBERT employs self-supervised learning to capture deep A-V contextual information, where the A-V features are masked and concatenated before fed into Transformer encoder to calculate masked-prediction loss for pre-training, and seq2seq loss is used for finetuning.
\item \textbf{u-HuBERT}~\cite{hsu2022u}: u-HuBERT extends the AV-HuBERT to a unified framework of audio-visual and audio-only pre-training.
\end{itemize}

\end{document}